\documentclass[a4paper,11pt]{article}

\usepackage[T1]{fontenc}
\usepackage{lmodern}
\usepackage[utf8]{inputenc}
\usepackage{authblk}
\usepackage[margin=1in]{geometry}
\usepackage{microtype}

\usepackage{tikz}
\usetikzlibrary{arrows.meta,positioning,shapes}

\usepackage{amssymb}
\usepackage{amsmath}
\usepackage{multirow}
\usepackage[hidelinks]{hyperref}
\usepackage{xurl}

\usepackage{listings}
\lstdefinestyle{erlangblock}{
   frame=single,
   language=Erlang,
   keywordstyle=\color{red},
   numbers=left,
   firstnumber=1,
   numberfirstline=true
}

\AtBeginEnvironment{lstlisting}{\lstset{style=erlangblock}}

\title{Erlang Binary and Source Code Obfuscation}
\author[1]{Gregory Morse\\\texttt{morse@inf.elte.hu}}
\author[1]{Tam\'{a}s Kozsik\\\texttt{kto@elte.hu}}
\affil[1]{E\"{o}tv\"{o}s Lor\'{a}nd Tudom\'{a}nyegyetem / University (ELTE), Budapest, Hungary}
\date{}

\begin{document}

\maketitle

\begin{abstract}
This paper studies obfuscation techniques for Erlang programs at the source,
abstract syntax tree, BEAM assembly, and BEAM bytecode levels.  We focus on
transformations that complicate reverse engineering, decompilation, and
recompilation while remaining grounded in the actual behavior of the Erlang
compiler, validator, loader, and virtual machine.  The paper categorizes
opcode-level dependency tricks, receive-based loop encodings, irregular
control-flow constructions, mutability-oriented performance obfuscation, and
self-modifying code enabled by dynamic module loading.  A recurring theme is
that effective obfuscation in BEAM often arises not from arbitrary corruption,
but from exploiting representational gaps between high-level Erlang semantics
and the lower-level execution model accepted by the toolchain and runtime.
\end{abstract}


\section{Introduction}

This paper focuses on low-level obfuscation techniques at the Erlang BEAM
bytecode level.  To keep the discussion accessible, we present high-level
Erlang source code wherever it faithfully conveys the same idea, and resort
to explicit BEAM instruction listings only when bytecode-level details are
essential to the obfuscation mechanism being described.

Obfuscation is any modification to compiled code which makes reverse
engineering it more difficult.  The goal of a reverse engineer may be to
either understand the code, modify the code, and ultimately recompile the
code.  Anything which makes any of those tasks increasingly complex falls
under the category of obfuscation.  In the software-protection literature,
this is typically framed as a semantics-preserving transformation whose goal
is to increase the cost of analysis, tampering, or reuse without changing the
observable behavior of the protected program
\cite{collberg1997taxonomy,xu2020layered}.

Binary obfuscation methods that we identified consist of: (1) BEAM-level
strategies which make use of inter-opcode dependencies; (2) structural and
control-flow based constructions which although 1:1 normally, can sometimes be
$n{:}1$; (3) \lstinline|try| with abnormal 1:$n$ and $n$:$n$ relationships
between usages of opcodes; (4) single try to multiple catch; (5) multiple try
to multiple catch; (6) multiple unnecessary \lstinline|wait_timeout|
instructions; and (7) multiple opening receive statements for a single receive
loop.  We further found techniques where loops can be constructed in BEAM via
the receive instruction with side effects.  We also found no-operation-like
binary sequences injected into code which can interfere with normal code
sequences, or nested conditionals, e.g.\ pattern matching.

We also found three source code obfuscation methods: (1) semantic information
derived from syntax tree; (2) semantic equivalence transformations; and (3)
return to syntax tree and source code.

The compiler is quite flexible but the BEAM validator will catch all sorts of
code that might theoretically still run without crashing the virtual machine;
however, it is designed with sufficient strictness to make sure no invalid
code is conceivable.  In particular, it is good at tracing errors like
uninitialized variables and heap allocations.  It is difficult to disable this
validation process.  It may require custom modification of the Erlang compiler
source code, or custom binary modification of actual \texttt{.BEAM} compiled
code.

Many BEAM instructions contain operand fields that are semantically inert for
program execution but are still checked by the BEAM validator.  Although such
arguments appear insignificant to a decompiler, they can be deliberately
manipulated to introduce misleading dependencies, disrupt static analysis, or
complicate reverse engineering without affecting runtime behavior.  The
remainder of this paper categorizes and illustrates such techniques at both
the BEAM and source-code levels.

From the reverse-engineering perspective, this paper sits at the intersection
of software obfuscation and decompilation.  The decompilation literature has
long treated control-flow recovery as one of the core obstacles in recovering
readable high-level programs from lower-level representations
\cite{cifuentes1995decompilation,schwartz2013phoenix,yakdan2015nomoregotos,gussoni2020comb}.
That perspective is especially relevant here because many of the obfuscation
strategies discussed below are designed specifically to frustrate those same
structuring and abstraction-recovery stages.

\paragraph{Contributions.}
The paper organizes the obfuscation space around transformation boundaries in
the Erlang toolchain, opcode-sequence dependencies, receive-driven loop
encodings, irregular BEAM control-flow shapes, performance-oriented mutable
tuple constructions, and dynamic self-modification through runtime
recompilation and module reloading.

Erlang is a particularly interesting setting for these questions because it
was designed for reliable distributed systems, lightweight processes, and
runtime code replacement, all of which make the boundary between source,
loaded code, and executing processes unusually rich from both an engineering
and reverse-engineering standpoint
\cite{armstrong2003reliable,armstrong2010erlang,erlang_code_loading,erlang_release_handling}.

\section{Transformation between BEAM, AST, BEAM Assembly, and Erlang Source}

Although BEAM itself and Erlang source have effectively one valid
representation, the AST has two versions, and the BEAM assembly, as a largely
internal representation, actually comes in three different varieties.  The
compiler produces \texttt{.S} assembly format from source \texttt{.erl} or AST
forms.  It then converts it into a tree-like structure for compilation via
\lstinline|compile:forms| which is ultimately the final entry into the
compiler.  However, the \lstinline|beam_disasm:file| method produces a
tree-like structure of the BEAM assembly that is incompatible with the
compilation library.  A flag \lstinline|DEBUG_DISASM| in that module will
allow, if it is recompiled, for five internal functions to be exported so that
\lstinline|beam_disasm:pp| can create the \texttt{.S} file.  To retrieve the
AST from a BEAM file that has debug information in it, which can be encrypted
with a password, \lstinline|beam_lib:chunks| is the necessary utility
function \cite{erlang_compile,erlang_beam_lib,erlang_absform}.

\paragraph{Note on Core Erlang.}
Core Erlang is an intermediate representation used internally by the Erlang
compiler between the abstract syntax tree (AST) and BEAM code generation.  We
do not include Core Erlang explicitly in the transformation graph for two
reasons.  First, Core Erlang is not exposed through stable public APIs in the
standard toolchain in the same way as AST forms, BEAM assembly, or BEAM
bytecode.  Second, the obfuscation techniques discussed in this paper operate
either above Core Erlang (at the AST or source level) or below it (at the
BEAM assembly and bytecode level), and therefore do not require direct
manipulation of Core Erlang itself \cite{erlang_cerl,erlang_compile}.

Nevertheless, Core Erlang remains an important conceptual stage in the
compilation pipeline, and its existence explains certain normalization and
simplification steps observed when translating from ASTs to BEAM code.

If desiring to get input valid for \lstinline|compile:forms| from Erlang
source, \lstinline|epp:parse_file| will suffice.  On the other hand, if one
has separate strings for the module definition, export definitions, and
function definitions as Erlang source code string snippets,
\lstinline|erl_scan:string| followed by \lstinline|erl_parse:parse_form| will
suffice.  Both of these strategies create an AST from Erlang source code.
However, if wishing to do this from a BEAM assembly \texttt{.S} source file,
then either the \lstinline|compile| module must be recompiled with
\lstinline|preprocess_asm_forms| exported, or the code for
\lstinline|preprocess_asm_forms|, \lstinline|collect_asm|, and
\lstinline|collect_asm_function| would have to be copied so that it provides
the appropriate translation \cite{erlang_epp,erlang_absform,erlang_compile}.

A graph summary of all the transformations and the functions that facilitate
them is shown in Figure~\ref{fig:beamtransform} to give a better visual
representation of the data flow.  In
general, it is not appropriate nor necessary to write any specific code to do
these conversions.  There are details like encoding format or other options
which affect compilation or validation and which must be taken into account.
One issue is that the compiling \texttt{.S} file format should have special
hints for the validator which will not be present when using the
\lstinline|beam_disasm| module, which will cause issues upon recompilation.
It is not trivial to provide these hints manually either.  However, validation
can be manually disabled by rebuilding the \lstinline|compile| module if such
functionality is needed \cite{erlang_compile,stenman2025beambook}.

\begin{figure}[ht!]
\centering
\resizebox{\textwidth}{!}{%
\begin{tikzpicture}[
  >=Stealth,
  box/.style={draw, rounded corners, align=center, font=\small, inner sep=3pt},
  line/.style={->, semithick}
]
  \node[box] (src) at (0,0) {.erl source file};
  \node[box] (ast) at (6,0) {AST v1/v2 format};
  \node[box] (compact) at (12,0) {compact BEAM asm tree};
  \node[box] (expanded) at (18,0) {expanded BEAM asm tree};
  \node[box] (asm) at (24,0) {.S BEAM asm file};
  \node[box] (beam) at (30,0) {.BEAM compiled file};

  \node[box] (pretty) at (3,-4) {erl\_syntax:form\_list $\rightarrow$\\ erl\_prettypr:format};
  \node[box] (compilefile) at (9,-4) {compile:file};
  \node[box] (preasm) at (15,-4) {compile:preprocess\_asm\_forms\\ compile:collect\_asm\\ compile:collect\_asm\_function};
  \node[box] (pp) at (21,-4) {beam\_disasm:pp\\ Need DEBUG\_DISASM};
  \node[box] (forms) at (27,-4) {compile:forms};

  \node[box] (epp) at (0,-8) {epp:parse\_file};
  \node[box] (strings) at (10,-8) {Erlang strings for\\ module, exports, functions};
  \node[box] (scanparse) at (18,-8) {erl\_scan:string $\rightarrow$\\ erl\_parse:parse\_form};
  \node[box] (chunks) at (26,-8) {beam\_lib:chunks\\ (with debug info)};
  \node[box] (disasm) at (34,-8) {beam\_disasm:file};

  \draw[line] (src) -- (compilefile);
  \draw[line] (compilefile) -- (preasm);
  \draw[line] (preasm) -- (compact);
  \draw[line] (compact) -- (forms);
  \draw[line] (forms) -- (asm);
  \draw[line] (forms) -- (beam);
  \draw[line] (asm) -- (compilefile);
  \draw[line] (beam) -- (chunks);
  \draw[line] (beam) -- (disasm);
  \draw[line] (chunks) -- (ast);
  \draw[line] (ast) -- (pretty);
  \draw[line] (pretty) -- (src);
  \draw[line] (disasm) -- (expanded);
  \draw[line] (expanded) -- (pp);
  \draw[line] (pp) -- (asm);
  \draw[line] (src) -- (epp);
  \draw[line] (epp) -- (forms);
  \draw[line] (strings) -- (scanparse);
  \draw[line] (scanparse) -- (forms);
\end{tikzpicture}%
}
\caption{Transformation graph between Erlang source, AST forms, BEAM assembly, and BEAM bytecode}
\label{fig:beamtransform}
\end{figure}

Since obfuscation requires a lot of ideas that step into the internals of the
BEAM assembly and specifics of VM execution, it is a prerequisite to be
familiar with all of these transformations.

\section{BEAM Assembly Details}

In BEAM, $x$ indicates registers where $x_0$ is used as the register
returning values from any function calls, while $y$ indicates local stack
variables which are explicitly allocated.  The heap is not explicitly accessed
but implicitly by instructions requiring it.  It is explicitly allocated by
\lstinline|allocate_heap| or \lstinline|test_heap| but implicitly
deallocated after instructions which use it.  This is largely transparent in
decompilation \cite{stenman2025beambook}.

\section{Custom Register Usage}

Unfortunately, as mentioned, if one tried to return values from a function in
$x_1$ instead of $x_0$, likely the emulator would still work correctly.  But
the BEAM validator will refuse to allow compilation to complete, as the called
function can return the value without issue, but the calling function cannot
make use of the value without triggering the validation failure.  Furthermore,
the virtual machine at its choice could dispose of any registers besides
$x_0$, so this is a VM-implementation-dependent concept.

Another idea is where an uninitialized value is used deliberately, perhaps as
part of entropy collection for a pseudo-random number generator.  Again, this
will not pass the validator.  The emulator as well could track these cases by
having very tight bounds on its cleanups and fail.

But the key here is that the concepts would yield valid BEAM code that may run
on the Erlang VM.  It is implementation dependent whether that holds or not.
Due to all the hints given in the BEAM assembly code regarding which registers
are still valid and being used, likely by properly adjusting and using these,
it would work.  In other words, register and stack liveness is not inferred
dynamically by the virtual machine.  Instead, the BEAM validator and loader
rely on the BEAM instruction stream itself to encode which registers remain
live and which may be safely discarded.  The VM executes under the assumption
that this information is correct, since it has no visibility into future
register or stack usage beyond what is explicitly described by the BEAM code.
This interplay between encoded invariants and low-level execution behavior is
part of what makes the BEAM a particularly interesting target for obfuscation
and reverse engineering \cite{stenman2025beambook,erlang_compile}.

As an example in Figure~\ref{fig:erlpatmatex}, having wanted to see what the
virtual machine actually stores when doing pattern matching on binaries, the
following module is compiled to BEAM.

\begin{figure}[ht!]
\centering
\begin{lstlisting}
-module(dumpbeam).
-compile(export_all).
dumpbinmatch() ->
	erlang:display({<<X, _, _>> = <<3, 4, 5>>}).
\end{lstlisting}
\caption{Erlang pattern matching example}
\label{fig:erlpatmatex}
\end{figure}

This yields the following BEAM assembly code in
Figure~\ref{fig:beampatmatex} when compiled with the ``S'' option, where we
could comment out the line which causes the result of the expression to be
returned after the match is already done since the equality operator returns
the left-hand side.

\begin{figure}[ht!]
\centering
\begin{lstlisting}
{function, dumpbinmatch, 0, 2}.
  {label,1}.
    {func_info,{atom,dumpbeam},{atom,dumpbinmatch},0}.
  {label,2}.
    {move,{literal,<<3,4,5>>},{x,0}}.
    {bs_start_match4,{atom,no_fail},0,{x,0},{x,0}}.
    {test,bs_test_tail2,{f,3},[{x,0},24]}.
    {move,{literal,{<<3,4,5>>}},{x,0}}.
    {call_ext_only,1,{extfunc,erlang,display,1}}.
  {label,3}.
    {badmatch,{literal,<<3,4,5>>}}.
\end{lstlisting}
\caption{BEAM code for the pattern matching example}
\label{fig:beampatmatex}
\end{figure}

But compiling \emph{after commenting out the line that returns the match
result} yields a validation error, shown in
Figure~\ref{fig:beampatmatvalerr}.

\begin{figure}[ht!]
\centering
\begin{lstlisting}
{error,[{"dumpbeam",
         [{beam_validator,{{dumpbeam,dumpbinmatch,0},
               {{call_ext_only,1,{extfunc,erlang,display,1}},
                9,
                {match_context,{x,0}}}}}]}],
       []}
\end{lstlisting}
\caption{Validation error after removing the result-preserving move}
\label{fig:beampatmatvalerr}
\end{figure}

So instead, we compile two versions of the module.  The first, renamed to
\texttt{.BEAM2}, contains the validator-compliant code shown in
Figure~\ref{fig:beampatmatvalid}.

\begin{figure}[ht!]
\centering
\begin{lstlisting}
    {move,{literal,{<<3,4,5>>}},{x,2}}.
    {move,{x,2},{x,0}}.
\end{lstlisting}
\caption{Validator-compliant idiom used as a patch source}
\label{fig:beampatmatvalid}
\end{figure}

The second contains the patched variant shown in
Figure~\ref{fig:beampatmatbinpatch}.

\begin{figure}[ht!]
\centering
\begin{lstlisting}
    {move,{literal,{<<3,4,5>>}},{x,0}}.
    {move,{x,0},{x,0}}.
\end{lstlisting}
\caption{Pattern-matching idiom prepared for binary patching}
\label{fig:beampatmatbinpatch}
\end{figure}

In fact, due to a bug in the VM BEAM loader (for which a pull request has
been submitted), the loader incorrectly optimizes away code sequences that were
not eliminable at compile time.  We describe this behavior not as a
recommended obfuscation strategy, but as a concrete example of how
post-validation BEAM patching can exploit gaps between validator assumptions
and VM execution.  Since the loader incorrectly optimizes out the code of the
first move statement as if it were dead code, and the second version already
works, calling \lstinline|dumpbeam:dumpbinmatch| yields
\lstinline|#MatchState|.  But doing a binary difference, with an Erlang
routine written to take a difference and patch, there will be two bytes,
followed by three double words, and finally a 16-byte block that are
differing.  The first two differences in position, BEAM2 value, and BEAM value
format are for example: \lstinline|[{411,3,35},{413,3,35},...]| and the $3$
value represents $x_0$ while $35$ or $23h$ represents $x_2$.  So we can patch
the value at $411$ in the BEAM file to change from $3$ to $23h$,
effectively changing $x_0$ to $x_2$, allowing $x_0$ to remain untouched and
thus bypassing validation.  Furthermore, we can change the value at $413$
from $3$ to $83h$, presumably $x_8$, and the output will contain ``[]''.

In general, uninitialized registers or local stack variables tend to have the
value ``[]'' which is considered to be ``nil'', but not as an atom in the
Erlang code sense, though it is an atom in the AST sense.  Therefore the
opcode \lstinline|test| with \lstinline|is_nil| would be expected to return
true.  It is the same as an empty list and is largely used as a placeholder
value.

This example illustrates the general theme of this paper: many BEAM-level
obfuscation techniques arise not from undefined behavior, but from precise,
implementation-dependent assumptions about validation, liveness, and
execution.

\section{Opcode Sequence Dependencies}

Initialization sequence opcodes for binaries by size of binary or bitstring
include \lstinline|bs_init2| and \lstinline|bs_init_bits|.  The only
difference between the opcodes is that the former has its size given in bytes,
while the latter uses bits.

Initialization sequence opcode for tuples by number of elements has only
\lstinline|put_tuple|.  Figure~\ref{fig:beamopseqdepex} gives an example of
this opcode in a dependency sequence.

\begin{figure}[ht!]
\centering
\begin{lstlisting}
{put_tuple,2,{x,0}}.
{put,{atom,ok}}.
{put,{x,1}}
\end{lstlisting}
\caption{BEAM opcode sequence dependency example}
\label{fig:beamopseqdepex}
\end{figure}

Element initialization opcodes for binaries include
\lstinline|bs_put_string|, \lstinline|bs_put_utf8|,
\lstinline|bs_put_utf16|, \lstinline|bs_put_utf32|,
\lstinline|bs_put_integer|, \lstinline|bs_put_binary|, and
\lstinline|bs_put_float|.  The string version takes a list of bytes, UTF-8 is
a character of 1, 2, 3, or 4 bytes depending on the value, UTF-16 is a
character consuming 2 or 4 bytes, while UTF-32 is fixed at 4 bytes.  The
integer, float, and binary versions can have their exact sizes specified as a
number of bits repeated some number of times, limited of course by the data
size, which for integer is not limited as it is a 32/64-bit
system-dependent integer or, if larger, a big integer; float is 64 bits,
while a binary, which could also be a bitstring, is of arbitrary size.

Element initialization opcode for tuples has only \lstinline|put|.

Constructing a binary from fixed data which are not constants will emit an
opcode which then requires a number of following opcodes which fill the binary
data to its initialized size.  Although the size will always have a numeric
value if compiled from Erlang, BEAM assembly allows any input, such as from a
register.  Therefore, a function could return the size, and a decompiler would
have no way, without being able to evaluate the expression precisely, to know
how many binary data-load instructions will follow.  Of course, heuristically,
it could just look for contiguous instructions which load the binary, but this
could be confused by intermixing other benign instructions in between such as
loading a register with an integer value.  The decompiler could instead opt to
store the latest binary which needs initialization so that any time a binary
load instruction occurs, it knows that it is the continuation point.  However,
control-flow could be introduced then to further increase difficulties for the
decompiler.  Since control-flow generally will not be cleaned up until a later
point, the decompiler will have a difficult time representing what is in fact
in the binary object.

The same is true for tuples constructed from fixed data.  It uses the same
initialization opcode followed by a sequence of opcodes which fill in the
elements of the tuple.  The tuple is more simplistic as the number of elements
does not require the same level of tracking as the size of the binary where
each element initialized could potentially contain a different size.

\section{Loops via Receive with Data Mutability}

\paragraph{Motivation and scope.}
The constructions in this section demonstrate that Erlang's
\lstinline|receive| mechanism can be used to encode loop structures that go
beyond the assumptions made by most Erlang compilers and decompilers.  In
particular, receive-based loops can simulate arbitrary control-flow graphs,
including multi-entry and multi-exit loops, by exploiting message passing,
mailbox state, and BEAM-level control-flow opcodes.

At the source-language level, this builds on Erlang's native message-passing
and process model, but it pushes those mechanisms into shapes that are far less
structured than the compiler or most reverse-engineering tools would normally
expect \cite{armstrong2003reliable,armstrong2010erlang}.

This significantly increases the complexity of control-flow analysis.  Only a
decompiler equipped with strong CFG normalization and structuring algorithms
can reliably recover recursive or iterative source-level constructs from such
code.  A decompiler targeting Erlang will almost certainly assume that these
patterns never occur, making them effective for obfuscation and stress-testing
analysis pipelines \cite{schwartz2013phoenix,yakdan2015nomoregotos,gussoni2020comb}.

To make these constructions concrete, we first summarize the BEAM opcodes
involved in receive-based looping and control flow.

Opcodes with control-flow back edges:
Unconditional: \lstinline|loop_rec_end|, \lstinline|wait|
Conditional: \lstinline|wait_timeout|

Entry to a receive block: \lstinline|loop_rec|
Exit and cleanup from receive block: \lstinline|remove_message|,
\lstinline|timeout|
Opcodes requiring $x_0$ to contain a valid message from the message queue:
\lstinline|loop_rec_end|, \lstinline|remove_message|

Conceptually we can convert a simple recursion example such as in
Figure~\ref{fig:erlsimprcrse}.

\begin{figure}[ht!]
\centering
\begin{lstlisting}
sum_to_n(0) -> 0;
sum_to_n(N) -> N + sum_to_n(N - 1).
\end{lstlisting}
\caption{Erlang simple recursion example}
\label{fig:erlsimprcrse}
\end{figure}

Into a receive-structured variant in Figure~\ref{fig:erlstrctrecv}.

\begin{figure}[ht!]
\centering
\begin{lstlisting}
sum_to_n_loop(N) ->
  U = erlang:unique_integer(), S = self(),
  Counter = 0, Sum = 0, S ! {U, N =:= 0},
  fun SumReceive(C, M, Sm) ->
	receive {U, X} ->
		if X -> Sm;
		  true -> P = M - 1, S ! {U, P =:= 0},
		      SumReceive(Counter+1, P, Sm+M)
		end
	  end
	end(Counter, N, Sum).
\end{lstlisting}
\caption{Erlang recursion via structured receive}
\label{fig:erlstrctrecv}
\end{figure}

The counter, although unused, will be needed to clean up the mailbox after the
loop ends which, although redundant and inefficient, is useful for
obfuscation.  It counts exactly how many times the loop body executed.  This
example demonstrates a for-loop or pre-test while loop.  A post-test while
loop would simply substitute the condition \lstinline|N =:= 0| for
\lstinline|false| for the first iteration.  The process itself is stored in a
variable to send messages, while a unique identifier is used to prevent any
interference; using the library unique integer function should be sufficient
for most cases, though otherwise a better unique identifier should be
generated.  The other variables are state variables for the loop.

Converting this to BEAM, on the other hand, requires a loop which replaces the
named fun expression in the Erlang listing.  Since it cannot actually remove
received messages, and due to technical details of messages which are not
being removed being considered ``fragile,'' meaning references to them cannot
be written to any state variables, and hence the validator preventing the
actual data in received messages from being used if not consumed, the message
itself cannot be used to store any state information beyond the unique ID and
the loop termination condition.  The other big difference is that after the
loop completes, all of the messages used to continue the loop will still
remain in the mailbox.  A very simple cleanup function can deal with them,
using the unique ID and the number of iterations through the loop body counter
in Figure~\ref{fig:erlmailuid}.

\begin{figure}[ht!]
\centering
\begin{lstlisting}
clear_uid_mailbox(_U, 0) -> ok;
clear_uid_mailbox(U, N) -> receive {U, _Any} ->
    clear_uid_mailbox(U, N-1) end.
\end{lstlisting}
\caption{Mailbox cleanup keyed by a unique identifier}
\label{fig:erlmailuid}
\end{figure}

This construction, although still virtual-machine dependent, is going to
largely work in all environments.  Partly this is because, as part of
validating the patterns for a given message in the queue, registers at a
minimum must be accessible, for example to extract items from tuples or lists
and compare them.  So in principle, the virtual machine has no idea or way of
restricting what changes in state occur between checking a message and either
accepting it by removing it or returning to the receive loop header to check
or wait for the next message.

This loop construct can be further extended for maximizing
``unstructuredness'' by using multiple exit targets from the loop.  In other
words, the control-flow at the different termination points in the loop can
jump further into the code in a way that no exit node post-dominates the loop.

Multiple entry or irreducible loops are also possible, but the
\lstinline|loop_rec| opcode must be used once for each entry and all except
the main entry must occur before entering the loop body.

These receive-based loop constructions illustrate how message passing and
mailbox manipulation can be used to deliberately defeat common Erlang
control-flow assumptions, reinforcing the need for robust, language-aware
structuring in decompilation.

While these constructions rely on BEAM execution semantics rather than
high-level Erlang syntax, they remain valid Erlang programs and therefore
represent a realistic adversarial input for decompilers.

\section{Irregular Control-Flow Constructions}

\paragraph{Motivation.}
This section documents BEAM-level control-flow constructions that are valid
with respect to the Erlang virtual machine but are either atypical or
impossible to express directly in Erlang source code.  The purpose is not to
suggest that such patterns arise in ordinary programs, but to demonstrate that
the BEAM execution model admits control-flow shapes, including multi-entry and
multi-exit regions, that violate the structural assumptions made by most
decompilers.  Understanding these constructions is therefore essential both
for designing robust control-flow recovery algorithms and for analyzing
obfuscation techniques that deliberately exploit such irregularities
\cite{cifuentes1995decompilation,schwartz2013phoenix,yakdan2015nomoregotos,gussoni2020comb}.

All of the constructs proposed are obfuscation by adding complexity to the
control-flow.  The compiler always optimizes the code to in fact take one-to-
one patterns and arrive at many-to-one or one-to-many variants to save space
and reduce the CFG.  The idea of using many-to-many relationships is to
thoroughly intermingle the CFG in a way that it causes cross edges that would
be difficult to resolve without a lot of code duplication, even an exponential
amount of code duplication.  Of course, the decompiler might afterward look
for patterns to undo such code duplication similar to the compiler
optimizations, and it could get back to the original source.  The decompiler
could alternatively use other techniques like those in the DREAM decompiler to
attempt to avoid code duplication but at the cost of adding conditions.  It
would still take code cleanup to arrive back at the original source.

The BEAM assembly instructions which implement \lstinline|receive|,
\lstinline|try|, and \lstinline|catch| have some peculiarities in their
control-flow that make awkward or abnormal implementations possible and
workable.  A decompiler which expects certain patterns would almost certainly
fail if control-flows not possible directly from the compiler were introduced.
The compiler does indeed do some rather clever optimizations which already
means a good decompiler needs to be fairly flexible in dealing with places in
the CFG where code paths merge.  Most of the idea here revolves around
creating multi-entry or multi-exit scenarios where opcodes expected to have a
one-to-one correspondence may have a many-to-one, one-to-many, or many-to-many
correspondence.  The compiler can in fact generate some of these.  Things are
further complicated by the fact that certain exception or termination
instructions allow cleanup opcodes to be omitted.

The \lstinline|receive| construct always has a one-to-one correspondence for
its various control-flow opcodes as well as its message handling body:
\lstinline|loop_rec|, \lstinline|loop_rec_end|, \lstinline|wait|,
\lstinline|wait_timeout|.  The jump or termination opcode
\lstinline|wait_timeout| is therefore accounted for, but termination opcode
\lstinline|remove_message| will appear once for each message pattern that is
matched.  Note that compiler optimizations are conceivable after the
termination opcode occurs.  But any many-to-many variants are possible here
based on arbitrary conditions.  Although to maintain semantic equivalence, it
would effectively be redundant control-flow or even duplication of the message
handling body, any type of complicated variants are possible.  However,
despite allowing arbitrary many-to-one or many-to-many control-flow variants,
the BEAM virtual machine enforces strict ordering constraints on
receive-related opcodes: \lstinline|loop_rec| must establish a receive context
before any matching or waiting occurs, and this context must be closed by
\lstinline|loop_rec_end| or \lstinline|wait_timeout| before message removal or
exit.  Violating this sequencing would break the VM's internal receive state
and is therefore not permitted.  This distinction---fixed opcode sequencing
versus flexible control-flow edges---is precisely what enables irregular yet
valid control-flow constructions.  For the control-flow path to be different,
some mutable data must be used.  This condition could come from the
non-matching message which is in $x_0$ between \lstinline|loop_rec| and
\lstinline|loop_rec_end|, or it could come from a random number generator,
use of some kind of mutable data in a stack register as per the section
discussing this.  A recommended technique is to call
\lstinline|erlang:process_info(self(), reductions).| as the reduction counter
constantly changes and has a relation to how Erlang processes work.  According
to the official Erlang/OTP documentation \cite{erlang_erts_manual}:
``This implementation-dependent function increments the reduction counter for
the calling process. In the Beam emulator, the reduction counter is normally
incremented by one for each function and BIF call. A context switch is forced
when the counter reaches the maximum number of reductions for a process (2000
reductions in Erlang/OTP R12B).''  Although the built-in functions (BIFs)
used, if any, during the loop would have to be taken into account.  Otherwise,
the function \lstinline|erlang:bump_reductions| which may be necessary anyway
in code dealing with mutable data in \lstinline|receive| can be used.
However, the documentation notes it may be deprecated.  But it is not clear
how this will be done when the \lstinline|recv_eval| handwritten assembly
module calls \lstinline|erlang:bump_reductions(134217727)| to implement the
\lstinline|eval| module.

The \lstinline|try| construct and its associated \lstinline|try_end| as well
as its related \lstinline|try_case| and \lstinline|try_case_end| can also
have any sort of many-to-many relationship.  Compiler optimizations will
already provide many-to-one relationships between these when it merges
duplicate code.

For simplification of a concrete example, we will use the \lstinline|catch|
and its associated \lstinline|catch_end| opcode.

In summary, these constructions illustrate that the BEAM instruction set
supports control-flow patterns far more general than those normally produced
by the compiler.  While such patterns are rarely encountered in practice, they
define the true semantic envelope within which a decompiler must operate.  Any
tool that assumes reducibility, single-entry regions, or fixed opcode
correspondence risks failure when confronted with either aggressive compiler
optimizations or intentional obfuscation
\cite{cifuentes1995decompilation,schwartz2013phoenix,yakdan2015nomoregotos,gussoni2020comb}.

\section{Efficiency via Mutability}

Another obfuscation tactic is to take data structures which are not efficiently
updatable with immutable code, as is the restriction with Erlang, and use
low-level BEAM to customize special routines for mutability.  Although still
decompilable, on recompilation it will, by design, be significantly slower
than an algorithm optimized with custom BEAM.  A decompiled version would
semantically represent the code correctly, but upon recompilation it will
cause a lot of copying and reconstruction which did not occur in the original
code.  The tuple is a very good specific example in this case.  It has a way
to set elements which according to the Erlang documents will only be used in
very specially identified cases.  Effectively this could be used as a very
fast efficient C-style array with custom BEAM code which would outperform any
of the various tree-structured objects like \lstinline|array| or linked-list
objects like \lstinline|lists| \cite{stenman2025beambook,armstrong2010erlang}.

This may not seem like obfuscation per se, but the fact that decompiled code
would not be reusable directly does represent obfuscation, not towards making
the code difficult to understand, but towards making it difficult to
recompile, as this routine would need to be disassembled and reassembled and
it would be very difficult for a decompiler to precisely understand these
mutable scenarios to output BEAM assembler and maybe a commented Erlang
decompilation of it.

Two examples are highlighted.  The first is a merge sort using tuples, which
will not outperform the highly efficient \lstinline|lists| implementation;
however, the second example, using random array reads and writes, shows that
it is highly performant.  A decompiled version however would need to retain
the BEAM assembly code and have the exact conditions for the tuple write
optimization checked or it would yield code that would be very slow.  There is
a key limitation in the two opcodes \lstinline|get_tuple_element| and
\lstinline|set_tuple_element|.  Both require constant integer values for the
indexing.  This would mean a separate function would be needed for each index
of the array, or specially generated functions for exact batches of setting or
getting values from the tuple.  The instruction for getting a tuple element is
probably so negligibly better that there is no point in even using it.  But
\lstinline|set_tuple_element| avoids memory copying.  Of course the
consequence is this violates immutability and it would actually modify the
original variables.  Of course for code which never uses old references, or is
designed for those old references to contain the modified value, it would be
safe and functional.  From the perspective of the Erlang programming model,
this is a completely unacceptable usage, but from a practical standpoint, it
will certainly work, it will be accepted by the virtual machine, and if the
code was designed properly have no side effects that are not well understood.

The compiler will refuse to generate \lstinline|set_tuple_element| without an
\lstinline|erlang:setelement| call preceding any sequence of them.  This is a
strict condition implemented though in fact a fake \lstinline|erlang:setelement|
call would trick the compiler as the check is not sophisticated but very
basic.  But it further can be overridden by copying the entire
\lstinline|compile| module source code and commenting out the following line
of the \lstinline|asm_passes| function:
\lstinline|?pass(beam_validator_weak),|.  Of course to load an alternative
\lstinline|compile| module which is built and marked as ``sticky'', one needs
to run: \lstinline|code:unstick_mod(compile).|  Then compilation and loading is
as simple as given in Figure~\ref{fig:erldynmodcompload}
\cite{erlang_compile,erlang_code_loading,stenman2025beambook}.

\begin{figure}[ht!]
\centering
\begin{lstlisting}
c("src/compile.erl", [debug_info,{outdir, "ebin"},
    {i,os:getenv("OTP_SRC_DIR") ++ "/lib/compiler/src"},
    {i,os:getenv("OTP_SRC_DIR") ++ "/lib/stdlib/include"}]).
\end{lstlisting}
\caption{Erlang dynamic module compilation and loading code}
\label{fig:erldynmodcompload}
\end{figure}

Where \lstinline|OTP_SRC_DIR| is an environment variable containing a valid
path to the root Erlang source code directory.  The Erlang developers are
adamant that an option to disable validation will never be possible.  They
specifically noted that if an object reference in such code from the offline
heap to the online heap were created, then garbage collection would crash the
emulator, such as in a case where a list on the heap is assigned to an element
of a tuple being used mutably.

First we need to be able to generate the relevant BEAM assembly functions in a
dynamically loaded module with no external files involved, as given in
Figure~\ref{fig:erlmuttuple}.

\begin{figure}[ht!]
\centering
\begin{lstlisting}
gen_put_asm(N) ->
  Names = list_to_tuple([list_to_atom("do" ++
      integer_to_list(X)) || X <- lists:seq(1, N)]),
  {ok, put_tuple_elem, CplBin} = compile:forms(
    {put_tuple_elem,[{element(X, Names), 2} ||
        X <- lists:seq(1, N)] ++
      [{module_info,0},{module_info,1}],[],
      [{function,element(X, Names),2,X*2,
          [{label,X*2-1},{func_info,{atom,put_tuple_elem},
        {atom,element(X, Names)},2},{label,X*2},
        {set_tuple_element,{x,1},{x,0},X-1},return]} ||
            X <- lists:seq(1, N)] ++
      [{function,module_info,0,N*2+2,[{label,N*2+1},
        {func_info,{atom,put_tuple_elem},
        {atom,module_info},0},{label,N*2+2},
        {move,{atom,put_tuple_elem},{x,0}},
        {call_ext_only,1,{extfunc,erlang,get_module_info,1}}]},
      {function,module_info,1,N*2+4,[{label,N*2+3},
          {func_info,{atom,put_tuple_elem},
        {atom,module_info},1},{label,N*2+4},{move,{x,0},{x,1}},
        {move,{atom,put_tuple_elem},{x,0}},
        {call_ext_only,2,{extfunc,erlang,get_module_info,2}}]}]
    ,7}, [binary, from_asm]),
  code:load_binary(put_tuple_elem, [], CplBin), Names.
\end{lstlisting}
\caption{Direct BEAM-assembly generation for mutable tuple updates}
\label{fig:erlmuttuple}
\end{figure}

This function will crash the virtual machine if $N > 524288$ or 512kb, which
is the maximum Erlang RunTime System (ERTS) export table size, making this
unfortunately not a simple solution as multiple modules would need to be used
or a specially compiled virtual machine with a greater limit, as it seems no
command line flags or other methods are possible to raise this limit which
appears to be hard-coded.

Here we present a merge sort algorithm using this strategy in
Figure~\ref{fig:erlmutmergesort}.

\begin{figure}[ht!]
\centering
\begin{lstlisting}
top_down_merge_sort(T) ->
  N = erlang:tuple_size(T), Names = gen_put_asm(N),
  B = setelement(1, T, element(1, T)),
  %list_to_tuple(tuple_to_list(T)),
  Result = element(1, top_down_merge_split(B, 1,
        erlang:tuple_size(T), T, Names)),
  code:purge(put_tuple_elem),
  code:delete(put_tuple_elem), Result.
top_down_merge_split(B, Begin, End, A, Names) ->
  if End - Begin < 1 -> {A, B};
  true -> Middle = (End + Begin) bsr 1,
    {BL, AL} = top_down_merge_split(
        A, Begin, Middle, B, Names),
    {BR, AR} = top_down_merge_split(
        AL, Middle+1, End, BL, Names),
    {BF, AF} = top_down_merge(BR, Middle, End, AR,
        Begin, Middle+1, Begin, Names), {AF, BF} end.
top_down_merge(A, Middle, End, B, I, J, K, Names) ->
  if K =< End ->
    if I =< Middle andalso (J > End orelse
       element(I, A) =< element(J, A)) ->
	  %setelement(K, B, element(I, A))
	  FuncName = element(K, Names),
	  top_down_merge(A, Middle, End,
	    put_tuple_elem:FuncName(
	      B, element(I, A)), I + 1, J, K + 1, Names);
    true -> %setelement(K, B, element(J, A))
	  FuncName = element(K, Names),
	  top_down_merge(A, Middle, End,
	    put_tuple_elem:FuncName(
	      B, element(J, A)), I, J + 1, K + 1, Names) end;
  true -> {A, B} end.
\end{lstlisting}
\caption{Erlang mutable merge sort}
\label{fig:erlmutmergesort}
\end{figure}

The comments show the likely output of a decompiler which must avoid BEAM
assembly code.  The merge sort $O(n \log n)$ will effectively have
$O(n^2 \log n)$ complexity if every placement of a value causes the entire
data structure to be copied.  This practically makes recompiled code unusably
slow for sufficiently large tuples.  Of course \lstinline|lists| has a highly
efficient algorithm for merge sort and even careful optimization as has been
done here cannot achieve its speed.  The measurement was performed without
considering the compilation time as there are easy ways to amortize such an
initialization.  However, the larger the size used, the slower
\lstinline|lists| gets in comparison.  But with the export table limitation,
it is unclear just how large the sorting object would need to be before
\lstinline|lists| becomes slower.  Details in copying, function call
invocation, and other practical issues are very likely making the difference
and the proper complexity is occurring in both variants.  These details will
of course eventually become insignificant at a large enough size.

So more appropriate might be checking how it compares with random access reads
and writes as if it were treated as an array in
Figure~\ref{fig:erltestmuttuple}.

\begin{figure}[ht!]
\centering
\begin{lstlisting}
test_random_read_write(ArrSize, Size) ->
  L = [rand:uniform() || X <- lists:seq(1, ArrSize)],
  T = list_to_tuple(L),
  Names = gen_put_asm(ArrSize),
  A = array:fix(array:from_list(L)),
  M = maps:from_list(lists:zip(lists:seq(1, ArrSize), L)),
  RR = random_reads(ArrSize, Size),
  RW = random_writes(ArrSize, Size),
  Result = {timer:tc(obfuscation, list_random_reads, [L, RR]),
    timer:tc(obfuscation, list_random_writes, [L, RW]),
    timer:tc(obfuscation, tuple_random_reads, [T, RR]),
    timer:tc(obfuscation, tuple_random_writes, [T, RW]),
	timer:tc(obfuscation, tuple_opt_random_writes,
	  [T, RW, Names]),
    timer:tc(obfuscation, array_random_reads, [A, RR]),
    timer:tc(obfuscation, array_random_writes, [A, RW]),
	timer:tc(obfuscation, maps_random_reads, [M, RR]),
    timer:tc(obfuscation, maps_random_writes, [M, RW])},
  code:purge(put_tuple_elem),
  code:delete(put_tuple_elem), Result.
random_reads(ArrSize, Size) ->
  [rand:uniform(ArrSize) || X <- lists:seq(1, Size)].
lists_setnth(1, [_|Rest], New) -> [New|Rest];
lists_setnth(I, [E|Rest], New) ->
  [E|lists_setnth(I-1, Rest, New)].
random_writes(ArrSize, Size) ->
  [{rand:uniform(ArrSize), rand:uniform()} ||
      X <- lists:seq(1, Size)].
list_random_reads(L, RR) -> [lists:nth(X, L) || X <- RR], true.
list_random_writes(L, RW) ->
  [lists_setnth(X, L, Y) || {X, Y} <- RW], true.
tuple_random_reads(L, RR) -> [element(X, L) || X <- RR], true.
tuple_random_writes(L, RW) ->
  [setelement(X, L, Y) || {X, Y} <- RW], true.
array_random_reads(L, RR) ->
  [array:get(X-1, L) || X <- RR], true.
array_random_writes(L, RW) ->
  [array:set(X-1, Y, L) || {X, Y} <- RW], true.
maps_random_reads(L, RR) ->
  [begin #{X := Y} = L, Y end || X <- RR], true.
maps_random_writes(L, RW) ->
  [L#{X => Y} || {X, Y} <- RW], true.
tuple_opt_random_writes(L, RW, Names) ->
  [put_tuple_elem:X(L, Y) || {X, Y} <-
     lists:map(fun({X, Y}) -> {element(X, Names), Y} end, RW)],
  true.
\end{lstlisting}
\caption{Random-access reads and writes with a mutable tuple}
\label{fig:erltestmuttuple}
\end{figure}

\lstinline|maps| unsurprisingly is clearly fastest as it has a very efficient
implementation and it has $O(1)$ read and write complexity.
\lstinline|lists| is unsurprisingly totally unusable in this context as
linked lists give $O(n)$ read and write complexity in the worst case.
\lstinline|array| reads are very fast but writes are slower, though both
should have theoretical worst-case complexity of $O(n \log n)$, except of
course that rebalancing the tree adds additional complexity.  The tuple
variant has $O(1)$ read complexity and $O(n)$ write complexity due to
copying, but the BEAM assembly variant reduces that to $O(1)$ complexity.  In
fact it appears to be faster than \lstinline|array|, which is interesting.
So now a full scenario where decompilation could potentially turn an $O(1)$
operation into an $O(n)$ operation is presented.  Of course there are several
other possibilities, not just with tuples but possibly with lists or binaries
as well.  Detailed knowledge of BEAM, algorithms, and the semantics of Erlang
and the BEAM virtual machine are all needed to attempt such ideas, however.

\section{Self-Modifying Code}

Although the BEAM format itself is clearly immutable after being loaded by the
emulator, Erlang has hot-swapping support where a module can be reloaded
dynamically when it is updated.  This is considered a key feature of the
Erlang language as it allows the system to continue operating when critical
code patches arrive.  Given its background in telecommunication systems, this
use-case fits the language particularly well.  However, this leads to some
interesting possibilities with self-modifying code.  As is typical in
functional programming languages, it would generally make sense to make the
self-modifications on the AST structure.  But it is also possible in direct
source code, especially when represented in the module, exports, and functions
string format.  As an example, one can partly use the strategy of writing a
proper quine in Erlang, where source code is able to print itself, but instead
do this at a function level so that instead of printing itself, it dynamically
reloads and calls itself.  This means a single function could potentially keep
modifying its own source code, reloading, and recalling itself
\cite{armstrong2003reliable,armstrong2010erlang,erlang_code_loading,erlang_release_handling}.

\begin{figure}[ht!]
\centering
\begin{lstlisting}
loop() ->
	io:format("Hello 1!~n", []),
	{ok, MTs, _} = erl_scan:string("-module(selfmod)."),
	{ok, ETs, _} = erl_scan:string("-export([loop/0])."),
	FH = "loop() -> io:format(\"Hello 2!~n\", [])," ++
	"{ok, MTs, _} = erl_scan:string(\"-module(selfmod).\")," ++
	"{ok, ETs, _} = erl_scan:string(\"-export([loop/0]).\")," ++
	"FH = \"~s\"," ++
	"{ok, FTs, _} = erl_scan:string(lists:flatten(io_lib:format(FH, [re:replace(FH, \"[\\\\\\\\\\\"]\", \"\\\\\\\\&\", [global, {return, list}])])))," ++
	"{ok, MF} = erl_parse:parse_form(MTs)," ++
	"{ok, EF} = erl_parse:parse_form(ETs)," ++
	"{ok, FF} = erl_parse:parse_form(FTs)," ++
	"{ok, selfmod, Bin} = compile:forms([MF, EF, FF], [binary])," ++
	"code:load_binary(selfmod, [], Bin)," ++
	"selfmod:loop().",
	{ok, FTs, _} = erl_scan:string(lists:flatten(io_lib:format(FH, [re:replace(FH, "[\\\\\\\"]", "\\\\&", [global, {return, list}])]))),
	{ok, MF} = erl_parse:parse_form(MTs),
	{ok, EF} = erl_parse:parse_form(ETs),
	{ok, FF} = erl_parse:parse_form(FTs),
	{ok, selfmod, Bin} = compile:forms([MF, EF, FF], [binary]),
  code:load_binary(selfmod, [], Bin),
  selfmod:loop()
.
\end{lstlisting}
\caption{Self-modifying Erlang function via dynamic recompilation}
\label{fig:erlselfmod}
\end{figure}

What makes this relevant for obfuscation is not merely that the program can
change itself, but that the representation boundary becomes fluid.  The same
logical routine can move between source strings, AST forms, compiled BEAM, and
runtime-loaded modules, making a purely static view of the program incomplete.
This is especially problematic for decompilers and reverse-engineering tools
which assume that code identity is stable once a module has been loaded.

In practice, such self-modifying techniques also provide a natural bridge to
staged obfuscation.  A program may begin from a relatively benign initial
representation, derive a transformed successor at runtime, load it into the
system, and continue execution in a form that was never present on disk in the
same way.  Even when the transformations are simple, this already complicates
analysis, replay, and reproducible recompilation.

\section{Conclusion}

The obfuscation techniques presented here show that Erlang's BEAM ecosystem
contains a much richer adversarial surface than would be suggested by source-
level Erlang alone.  Transformation boundaries between source, AST, assembly,
and bytecode create opportunities for representational mismatch; validator and
loader assumptions create opportunities for subtle binary-level manipulation;
receive, exception, and mutable tuple mechanisms create opportunities for
unusual control-flow and performance-sensitive constructions; and dynamic
loading enables self-modifying behavior that resists purely static analysis.

From the perspective of reverse engineering, the central lesson is that a
faithful Erlang decompiler cannot merely reconstruct source-level syntax.  It
must instead reason about the semantics and constraints of the BEAM validator,
loader, mailbox model, and runtime system.  From the perspective of
obfuscation, the same lesson suggests that the most effective transformations
are often those that remain semantically disciplined while exploiting the gap
between what high-level tools assume and what the BEAM execution model actually
permits.

\bibliographystyle{plain}
\bibliography{biblio}

\end{document}